\documentstyle[aps,prl,twocolumn]{revtex}
\input epsf
\begin{document}
\draft
\title{Oscillatory Driving of Crystal Surfaces: a Route to Controlled
Pattern Formation}
\author{O.\ Pierre-Louis$^*$ and M. Haftel$^\dagger$}
\address{$^*$LSP-GREPHE, CNRS, UJF-Grenoble 1,
BP87, F38402 Saint Martin d'H\`eres, France.\\
$^\dagger$ Nanostructure Optics Section, Naval Research Laboratory, 
Washington DC 20375-5343, USA.}
\date{\today}

\maketitle
\begin{abstract}

 We show that the oscillatory driving 
of crystal surfaces can induce pattern formation or
smoothening. The driving force can be 
of quite different origin such as a pulsed laser beam, an
electric field, or elasticity. Depending on driving conditions,
step bunching and meandering, 
mound formation, or surface
smoothening may be seen in presence of a kinetic asymmetry 
at the steps or kinks (the
Ehrlich-Schwoebel effect). We employ a step model to calculate the
induced mass flux along misoriented surfaces, 
which accounts for surface dynamics and stability. 
Flux inversion is found when varying the driving frequency. 
Slope selection and metastability result from the
cancellation of the mass flux along special orientations. 
Kinetic Monte Carlo simulations illustrate
these points.
\end{abstract}
\vspace{0.3in}
\vspace{-0.3cm}

Sculpting surfaces at nanometer scale is
of major technological interest. Besides lithography,
spontaneous pattern formation during crystal growth has been proposed
as a tool to create large scale nanostructured surfaces.
In this letter we point out that 
oscillatory driving appears as an alternative route for pattern formation. 
We see two basic advantages in this method:
first patterning and growth are separated, 
so that morphology is not a function of the growth process.
Second it offers better control of the structure. An
in situ and real-time control of the pattern becomes possible, opening
a wide range of new applications.

On the side of fundamental physics, 
oscillatory forcing has been a long standing
source of intriguing nonlinear phenomena,
such as the inverted pendulum problem \cite{kapitza65},
and the Faraday instability \cite{faraday1831} in fluid mechanics.
Oscillatory forcing also induces pattern formation
in granular media \cite{sanddish}.
Moreover, parametric forcing of an ensemble of oscillators
is known to lead, for example, to motion of domain walls \cite{coullet90}.
In this letter, we show that oscillatory forcing of crystal surfaces
induces macroscopic mass fluxes, leading to
pattern formation or smoothening.

In the past 15 years, a large number of studies were devoted to
surface roughening during crystal growth.
We show that all the instabilities identified
during growth (namely mound formation, step meandering, and step bunching)
appear under oscillatory driving.
The physical origin of this effect can be related to ratchets,
in the sense that kinetic anisotropy of steps (Erhlich-Schwoebel (ES) effect)
is used to produce a mass flux along misoriented
surfaces, as pointed out by Barabasi {\it et al} \cite{bara98},
who showed that AC electromigration
should lead to directional smoothening of surfaces.

We first derive the mass flux 
along a misoriented (vicinal) surface
in order to analyse the stability of the surface.
The main results are illustrated by Kinetic Monte Carlo simulations.
We also briefly mention
some experimental situations where our analysis applies,
such as pulsed laser-induced 
mound formation on metal-vacuum surfaces, 
ultra-sound driven pattern formation on thin films \cite{tucou99},
and pattern formation at metal-electrolyte interfaces
subject to an oscillating electrochemical potential.

We consider surfaces where adsorption, desortion, and 
defect creation (such as bulk vacancies) are not allowed.
Mass transport then only occurs through surface diffusion.
Since the mean height of the surface does not
vary with time, we have
\begin{eqnarray}
\partial_t\langle h\rangle=-\nabla . \langle{\bf  j}\rangle \; ,
\label{e:aver_mass_cons}
\end{eqnarray}
where the brakets indicate that we have averaged over
the timescales of oscillations,
and $\partial_t\equiv \partial /\partial t$.
The mass flux $\langle {\bf j}\rangle$ is the key quantity 
for surface dynamics.
Step properties are supposed to be isotropic
and direct step interactions via elasticity, or electronic surface
states, are neglected. Steps then only interact via mass exchange.
The surface flux has two components 
$\langle{\bf  j}\rangle=\langle J \rangle {\bf n}+\langle G \rangle {\bf s}$
where ${\bf n}=\nabla h/|\nabla h|$ and ${\bf s}.{\bf n}=0$.
From Eq.(\ref{e:aver_mass_cons}) it follows that a nominal surface is stable
(unstable) if $\langle J\rangle>0$  ($<0$). 
On a vicinal surface of mean slope $m_0\neq 0$, 
an uphill flux $\langle J\rangle>0$ also brings a destabilizing contribution
to step meandering (i.e. undulations perpendicular to the 
mean slope direction). Nevertheless, in this case, the $\langle G \rangle$
component also intervenes in the stability criterion.
A vicinal surface of mean slope $m=|\nabla h|/a$, where $a$ 
is the lattice spacing, is stable
with respect to step bunching (i.e undulation
along the mean slope) if $\partial_m\langle J\rangle>0$.
Otherwise it is unstable.

\begin{figure}
   \centerline{\epsfysize=5cm\epsfbox{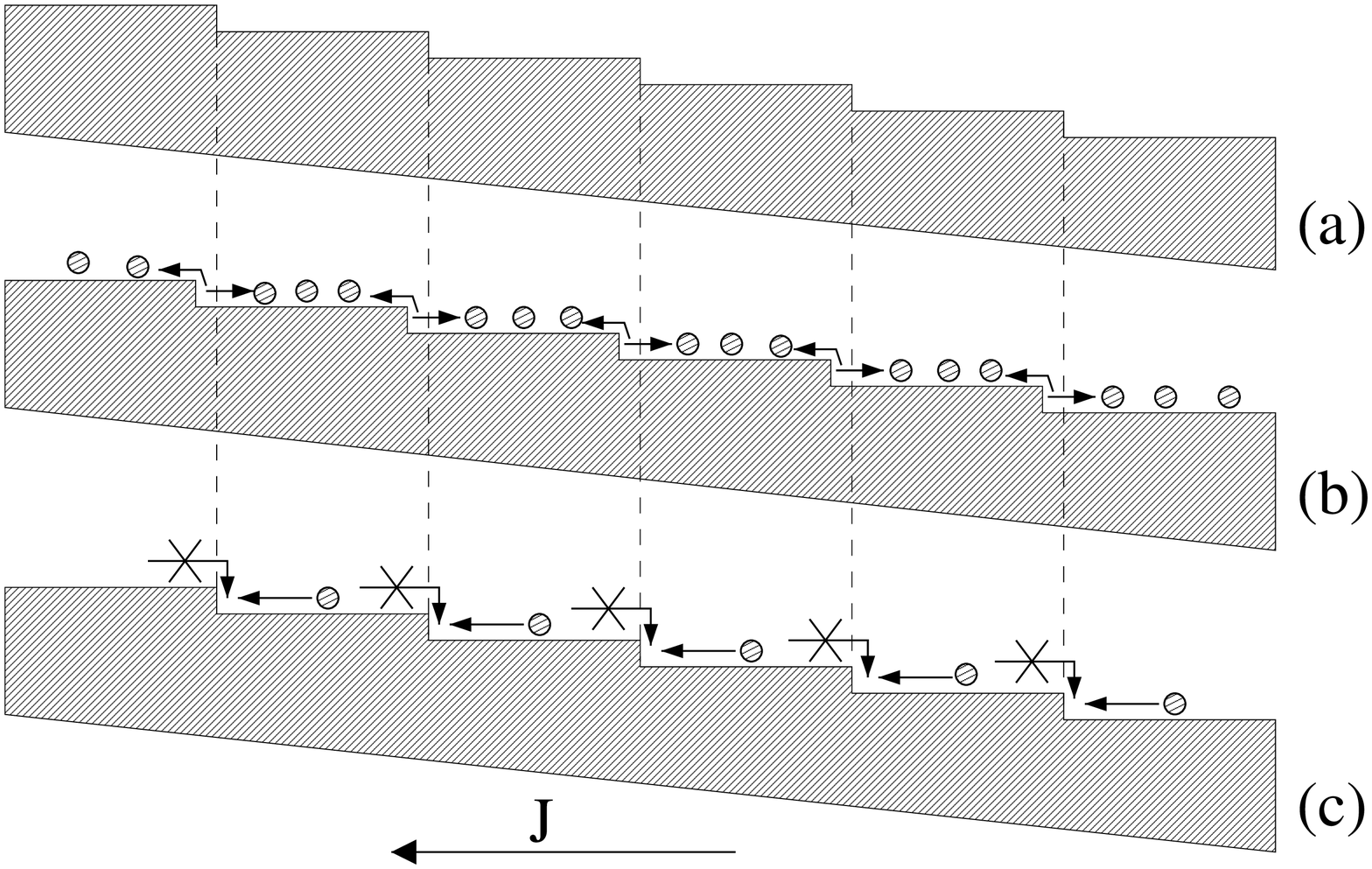}}
   \caption{
Mechanism for an uphill flux, see text.}
\label{fig1}
\end{figure}

As a first qualitative example of a mass flux induced
by oscillatory driving, let us consider a vicinal surface
for which the temperature oscillates alternatively  and
abruptly between two values. In the high temperature regime,
the equilibrium concentration on terraces is high, and
the ES effect is repressed. In the low temperature
regime, the equilibrium concentration is low,
and the ES effect is strong (i.e. adatoms cannot
attach going down-step). We start from a regular vicinal surface
(Fig. \ref{fig1} (a)) in the low temperature regime.
We first switch to  the high temperature regime:
atoms detach from the steps and go on terraces 
(Fig. \ref{fig1} (b)). Steps slightly
retract to the left.
Switching back to low temperature,
adatoms go back to the steps but only attach from ahead
(Fig. \ref{fig1} (c)).
The step goes back to its initial position.
A net uphill flux results from the last part of the cycle
(This could be interpreted as a many-particle "ratchet effect").

In the following we calculate the mass flux $\langle J \rangle$
along a misoriented (vicinal) surface, where steps
are regularly spaced (with a distance $\ell=1/m$).
Dynamics of a vicinal surface are decribed by 
the 1D step model of Burton Cabrera and Frank \cite{bcf51},
modified by Schwoebel \cite{schwoebel69} in order to  account for
step kinetics. On terraces between steps,
the adatom concentration only evolves via diffusion:
\begin{eqnarray}
\partial_t c=D\partial_{xx}c \, ,
\label{e:diff}
\end{eqnarray}
where $D$ is the adatom diffusion constant and $\partial_t$ denotes
a time derivative. 
At the steps, mass conservation reads:
\begin{eqnarray}
{\partial_t z \over \Omega}=D\partial_xc_+-D\partial_xc_-
\label{e:mass_cons}
\end{eqnarray}
where $\Omega$ is the atomic area,
$z$ denotes the step position, and the index $+$ and $-$
indicates the low and high sides of the step respectively. 
Relation (\ref{e:mass_cons})
is valid when $\Omega c_\pm \ll 1$, i.e. 
when the adatom concentration is much smaller than that of the solid.
In order to describe attachment and detachment kinetics at the steps,
the incoming diffusion flux is related to departure from equilibrium
\cite{schwoebel69}:
\begin{eqnarray}
D \partial_xc_\pm=\pm\nu_\pm (c_\pm-c_{eq})
\label{e:step_kin}
\end{eqnarray}
where, $\nu_\pm$ are kinetic attachment coefficients.
We define the kinetic attachment lengths $d_\pm=D/\nu_\pm$,
that are small for fast kinetics and large for slow kinetics.

Let us first consider
a sinusoidal pertubation:
\begin{eqnarray}
D&=&D_0+ D_1 \cos(\omega_0t)\,
\nonumber \\
d_\pm&=&d_{0\pm}+d_{1\pm}\cos(\omega_0t)\, ,
\nonumber \\
c_{eq}&=&c_{0eq}+ c_{1eq}\cos(\omega_0t)
\label{e:param}
\end{eqnarray}
where quantities with index $1$ are small and not necessarily
positive.
The mean flux going through a step is
\begin{eqnarray}
\langle J \rangle= -{D \over 2}
\langle \partial_xc_++\partial_xc_- \rangle
\label{e:flux_gen}
\end{eqnarray}
To zeroth order in the perturbation the solution of 
Eq. (\ref{e:diff}) with the boundary condition (\ref{e:step_kin}) is 
$c=c^0_{eq}$  yielding zero contribution to $\langle J\rangle$. 
The oscillatory nature of the perturbation yields zero
first-order contribution.
From the second order solution of Eqs. (\ref{e:diff},\ref{e:step_kin}) 
we obtain the mean flux:
\begin{eqnarray}
\langle J_2 \rangle=
{\Omega \over 2}
{D_0 c_{1eq}m \over 1+m(d_{0+}+d_{0-})}
\times \nonumber \\
\Re e\left[ \lambda
{d_{1+}({\rm ch}-1+\lambda d_{0-}{\rm sh})
-d_{1-}({\rm ch}-1+\lambda d_{0+}{\rm sh})
\over
(1+d_{0+}d_{0-}\lambda^2){\rm sh}
+\lambda(d_{0-}+d_{0+}){\rm ch}}
\right]
\label{e:j2}
\end{eqnarray}
where ${\rm ch}=\cosh(\lambda/m)$ and ${\rm sh}=\sinh(\lambda/m)$,
and $\lambda^2=i\omega_0/D$.
The mean flux $\langle J_2 \rangle$ results from a combination
of oscillations of the equilibrium concentration and step kinetics.
The frequency and slope dependence of $\langle J_2 \rangle$ is in general
complicated. 
We do not wish to be exhaustive here,
but rather to highlight some important features.

\begin{figure}
   \centerline{\epsfysize=3cm\epsfbox{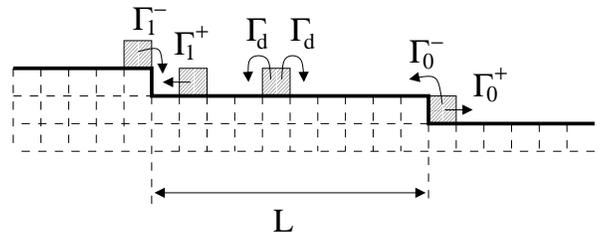}}
   \caption{
Hopping frequencies in a discrete model}
\label{fig2}
\end{figure}

In the high frequency limit, where $\omega_0$ is
smaller than phonon frequencies $\omega_p\sim 10^{12}s^{-1}$,
(in order to avoid effects such as stochastic resonance
with phonons), but {\it larger than atom hopping frequencies} $\Gamma$,
one can question the validity of Eq. (\ref{e:j2}), 
which is based on a step model (Eq.(\ref{e:diff}-\ref{e:step_kin})),
in the "hydrodynamic limit" (i.e. low frequency). 
Typically, the slowest rate is that of detachment from steps
$\Gamma_0\approx 10^{12}{\rm exp}(-E_0/k_BT)$. Taking typical values
$E_0\sim 0.5 eV$, $T\sim 300K$, one obtains $\Gamma_0\sim 10^3s^{-1}$.
We use a discrete 1D model to check the high frequency limit
$\omega_p \gg \omega_0 \gg \Gamma$. 
In this model, atoms hop
to nearest neighbors --without interacting, 
on a frozen periodic vicinal surface of interstep distance $\ell=La$.
Dynamics is described with help of hopping frequencies $\Gamma_d$,
$\Gamma_0^{\pm}$, and $\Gamma_1^{\pm}$ for diffusion,
detachment and attachment on both sides of the step site 
(see Fig. \ref{fig2}). Hopping frequencies can be related to
the parameters of the step model via
$D=\Omega\Gamma_d$, $d_\pm=a(\Gamma_1^{\pm}/\Gamma_d-1)$ \cite{politi96},
and $\Omega c_{eq}=\Gamma_0^{\pm}/\Gamma_1^{\pm}$.
These relations can be inverted to express frequencies
as a function of the BCF model parameters. 
Using Eq.(\ref{e:param}) and in the $\omega_0 \rightarrow \infty$ limit,
average frequencies can be used for atom hops.
One then finds that detailed balance
is broken to second order in the perturbation: 
a steady mass flux along the surface is present. 
Since we considered a "tracer" model on a frozen
surface, an additional condition is necessary to determine 
this flux, because the amount of matter on terraces is not fixed yet.
We choose the coverage at the step site to be $\theta_0=1$
( i.e. there is always an atom at the edge of a step).
We then find the following steady flux:
\begin{eqnarray}
J_{\rm steady}= {D_0 \Omega c_{1eq}m \over 2(1+m(d_{0+}+d_{0-}))}
\left({d_{1+} \over a+d_{0+}}-{d_{1-} \over a+d_{0-}}\right)
\label{e:high_freq}
\end{eqnarray}
Taking the limit $\omega_0 \rightarrow \infty$ in Eq. (\ref{e:j2})
leads to the same result, except that the $a$'s in the denominator
are absent. This suggests that the step model is valid in the high frequency
limit as long as step kinetics are not too fast i.e. $d_{0\pm}>a$. 

In the low frequency limit $\omega_0 \rightarrow 0$, the flux vanishes:
as $\langle J_2 \rangle_{\infty} \sim \omega_0^2$.
In this limit, a wide variety of slope dependances can be obtained. 
We shall first focus on the occurence of {\it flux inversion} 
as frequency varies.
When attachment-detachment is fast, i.e. when
$md_{0\pm}\ll 1$, (and for $\omega_0\rightarrow 0$), 
an expansion of (\ref{e:j2}) provides:
\begin{eqnarray}
\langle J_2 \rangle={\Omega c_{1eq} \over 48D_0} {\omega_0^2 \over m^2}
(d_{1+}-d_{1-})
\label{e:low_freq_wese}
\end{eqnarray}
By comparison to Eq. (\ref{e:high_freq}),
it is seen that if
$(d_{1-}/d_{0-}-d_{1+}/d_{0+})(d_{1-}-d_{1+})<0$,
there must exist a frequency $\omega^*$ for which 
$\langle J_2\rangle$ changes sign. Hence, surface
stability can be changed by tuning the frequency. 

In the limit of slow step kinetics $md_{0\pm}\gg 1$
(and $\omega_0\rightarrow 0$),
we find:
\begin{eqnarray}
\langle J_2 \rangle={\Omega c_{1eq} \over 2D_0} {\omega_0^2 \over m^2}
\left({d_{1+} \over d_{0+}}-{d_{1-} \over d_{0-}}\right)
{d_{0+}^2d_{0-}^2 \over (d_{0+}+d_{0-})^3}
\label{e:low_freq_sese}
\end{eqnarray}
Comparing the sign of this expression to that of Eq.(\ref{e:low_freq_wese}),
we see that $\langle j_2 \rangle$ might change for a special slope $m^*$.
Two cases are possible:
(i) If $d_{1+}<d_{1-}$, and $d_{1+}d_{0-}>d_{1-}d_{0+}$,
then a nominal surface is unstable, but the flux changes sign
for $m=m^*$. In this situation, {\it slope selection} is expected
\cite{siegert94}. 
(ii) If $d_{1+}>d_{1-}$, and $d_{1+}d_{0-}<d_{1-}d_{0+}$,
the surface is stable with respect to small fluctuations,
but mounds of slopes larger than a special slope $m^*$ 
are subject to an uphill mass flux. 
A surface of finite initial roughness may be destabilized although 
a flat one is linearly stable: it is {\it metastable}.
An analogous metastable situation occurs for morphologically
unstable steps in presence of kink Ehrlich-Schwoebel effect during growth
\cite{oplmrdtle}. 

Since mass fluxes are second order,
they strongly depend on the temporal ``shape" of the 
perturbation. As an example consider 
the low frequency square pertubation mentionned earlier (Cf. Fig.\ref{fig1}).
Here $\cos(\omega_0t)$ is replaced
by ${\rm sign}[\cos(\omega_0t)]$ in Eq. (\ref{e:param}),
leading to the following flux, calculated
for finite perturbation amplitudes
(i.e. $d_{\pm 1}$ and $c_{eq}^1$ are not small):
\begin{eqnarray}
\langle J \rangle \approx 
{\Omega c_{eq}^1\omega_0 \over 4\pi}\; 
{d_{1+}-d_{1-} + 2 m(d_{1+}d_{0-}-d_{1-}d_{0+})
\over [1+m(d_{0}+d_{1})] [1+m(d_{0}-d_{1})]}
\label{e:square_flux}
\end{eqnarray}
where $d_0=d_{0+}+d_{0-}$ and $d_1=d_{1+}+d_{1-}$.
The slope and  frequency dependance
of this flux are different from the sinusoidal case [Eq.(\ref{e:j2})],
leading to new conditions for surface stability.
Taking $d_{0+}=d_{1+}=0$, $c_{1eq}=c_{0eq}$ and $d_{1-}=d_{0-}$ (case A),
which corresponds to the case depicted in Fig \ref{fig1}, we obtain
an uphill flux, leading to mound formation and step meandering. 
If instead $d_{1-}=-d_{0-}$, a downhill flux is found
(case B). Nominal surfaces should then be smoothened
and step bunching is expected on vicinal surfaces.

\begin{figure}
   \centerline{\epsfysize=8cm\epsfbox{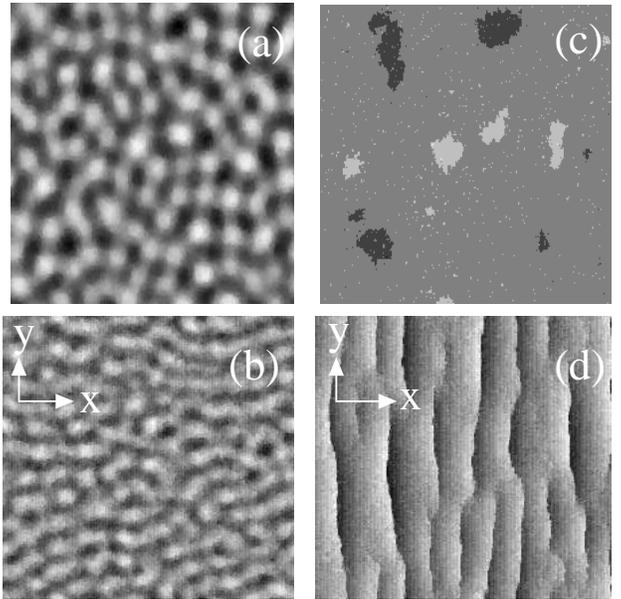}}
\caption{SOS simulations on a $256 \times 256$ lattice,
greyscale represents surface height.
$\omega_0/2\pi=0.1$ MCSPS$^{-1}$, $E_b=1.$, $E_{1b}=0.3$,
$E_s=2$, and $T=0.4$. For (a) and (b) $E_{1s}=1.9$;
for (a) and (d) $E_{1s}=-1.9$. Starting from a flat 
nominal surface (a) is otbained after $5\times 10^{5}$ MCSPS.
Mound formation is seen and r.m.s. roughness is $w\approx 8$. 
Starting from the patterned
surface (a), we obtain (c) after $2\times 10^5$ MCSPS,
with $w\approx 0.25$.
In (b) and (d) we have plotted the height
minus that of the intial regular vicinal surface.
Steps are initally parallel to the $y$ axis.
Meandering (b) and bunching (d) of steps take the form of ripples
along $x$ or $y$.}
\label{fig3}
\end{figure}

We have performed
Kinetic Monte Carlo simulations in order to show that
the above mentionned analysis allows one to predict
the surface response to an oscillatory 
variation of hopping rates.
We use a simple Solid on Solid (SOS) model on a square lattice, 
where hops are accepted with an Arrhenius law,
the activation energy being equal to the bond energy $E_b$ times
the number of in-plane nearest neighbors. 
An extra energy barrier $E_s$, accounting
for the Ehrlich-Schwoebel effect, is experienced
during interlayer hops.
Taking the lattice constant
as the unit length, and one Monte Carlo Step per site
(MCSPS) as the unit time, we have $D=1/4$, $d_+=0$
(i.e. $d_{0+}=d_{0-}=0$), $d_-={\rm exp}[E_s/T]-1$.
$E_b$ and $E_s$ oscillate in a square
fashion, between $E_{0b}-E_{1b}$ and $E_{0b}+E_{1b}$
(and similarily for $E_s$) leading to case A or B
depending on wether their oscillations are in-phase, or out-of-phase. 
We use the following parameters $E_b=1.$, $E_{1b}=0.3$,
$E_s=2$, $E_{1s}=\pm 1.9$, $T=0.4$, and $\omega_0/2\pi=0.1$. 
The resulting patterning of the surface correspond to
the expectations, as seen in Fig. \ref{fig3}.

We also studied
the coarsening of these structures on a nominal surface.
We have performed simulations
for switching frequencies $1/10$, $1/30$, and $1/100$. In all cases,
surface roughness behaves as $t^\beta$, with $\beta \approx 0.44\pm 0.02$.
The typical lateral length-scale on nominal surfaces also follows 
a scaling law $t^\alpha$,
with $\alpha=0.09\pm0.02$.
The up-down asymmetry of the mounds is smaller than during growth.
This might be related to the non-zero mean step velocity
during growth \cite{politi96}.
As expected, the surface can also be smoothened
(see Fig.3(c)).

Let us now turn to some applications of this phenomenon.
A pulsed laser could induce the abrupt
temperature change  leading to the uphill flux
presented in Fig. \ref{fig1}. 
Conditions similar to that of Ref. \cite{ernst98} would be needed.
For our analysis to be valid, the surface should not melt, 
and no dislocation should be induced.

In recent experiments, high amplitude surface ultra sound waves
were applied to an Al thin film  \cite{tucou99}, 
and pattern formation was observed. It is not clear
wether this results from the oscillatory driving or from Grinfeld
instability as pointed out in Ref. \cite{tucou99}. 
Moreover, further experiments would be needed
in order to determine wether dislocations are present.
 
The oscillatory driving of the potential in an electrochemical cell can 
provide morphological changes. On the basis of the 
surface-embedded-atom-method (SEAM) \cite{hr95}
we have surveyed the dependence of 
the ES barrier and the adatom equilibrium concentration on Ag(111) as a 
function of the departure of the electrochemical potential from the 
potential of zero charge (pzc) \cite{he99}. 
We find that increasing the potential 
from 0 to +0.85 V (relative to the pzc) increases
the adatom formation energy from $0.85$eV to $0.92$eV
(which implies a decrease of the adatom 
equilibrium concentration), but also increases the ES 
barrier from $0.22$eV to $1.16$eV. This cycle thus should lead to case A.
We also find that as the potential is cycled from 0 to 
negative voltages, $c_{eq}$ still increases, but the ES barrier goes back 
up. Therefore cycling between $0$V and negative voltages 
we should be in case B.
Hence, it should be possible to create or smoothen mounds
depending on the chosen cycle.

Note that oscillatory driving could also be applied
during growth via the techniques described above, in order
to smoothen or pattern the surface. 

In conclusion, oscillatory driving is a promising tool 
for in situ and real time control of the surface morphology.
We have shown that it allows one to create and smoothen
patterns on nominal and vicinal surfaces.
A wide variety of behavior among which step bunching and meandering,
mound formation, slope selection, metastability,
and mass flux inversion are found.
Similar phenomena should result from a Kink-Schwoebel effect,
as a consequence of non-equilibrium line diffusion, as
shown for growth in Ref. \onlinecite{oplmrdtle}.

In order to be more quantitative and to predict
the typical wavelength of the instabilities, 
the next step of this study will be to consider stabilizing effects 
coming from line tension, step interactions, and nucleation.
As in growth, initial stages of the instability on nominal surfaces,
as well as the shape of mounds crucially depend on nucleation.
Including this effect is the main challenge for a global understanding
of pattern formation via oscillatory driving.

Aknowledgements: The authors wish to thank  
T.L. Einstein, C.Caroli, J.Lajzerowicz, T. Biben, and C. Misbah
for useful discussions. The authors appreciate the hospitality of the Univ. 
of Maryland MRSEC, where part of this work was carried out. Funding of the 
Office of Naval Research and support by the Department of Defense
High Performance Computation Modernization Project is aknowledged.

\end{document}